\documentstyle[12pt]{article}
\input tcilatex
\begin{document}
\centerline{\large ASYMPTOTICS OF SOLUTIONS OF}\vskip.2cm
\centerline{\large THE DISCRETE STRING EQUATION}\vskip.5cm
\centerline{Vadim L. Vereschagin \footnote[1]{supported by the ISF grant RK2000
and RFFI grant 94-01-01308}}\vskip.2cm
\centerline{Irkutsk Computing Center, P.O.Box 1233, Irkutsk 664033, RUSSIA}

\vskip 1.0cm

\par {\bf Abstract.} The main subject of the paper is the so-called
Discrete Painlev\'e-1 Equation (DP1). Solutions of
DP1 are classified under criterion of their behavior while argument tends to
infinity. The Isomonodromic Deformations Method yields asymptotic formulae for
regular solutions of DP1. DP1 is an integrable system, what allows to develop
appropriate Whitham theory. Asymptotics of singular solutions of DP1 are
calculated by using the Whitham method.

\vskip .5cm\par
{\bf Key words.} Painlev\'e eqs., asymptotics, isomonodromy method
\vskip .5cm \par

\par {\bf Introduction.} Recent rush of interest to ordinary differential
equations of the Painlev\'e type (see \cite{1} ) is provided mostly by two
reasons.
The first one consists in rich history of the Painlev\'e equations as a
classical object of the ODE analytical theory. The second one is that they
arise in a number of concrete problems in different aspects of theoretical and
mathematical physics related to nonlinear evolutionary equations, quantum
field theory and statistical physics (see \cite{2,3,4,5} ). Analysis of some
models of
nonlinear theoretical physics indicates that the Painlev\'e transcendents,
describing self-similar regimes, play a role very similar to that of classical
special functions in linear problems. So, the first Painlev\'e eq. (P1)

$$y''=6y^2+x\eqno(1)$$
and the second one

$$y''=xy+2y^3+\nu_{2},\ \ (\nu_{2} \ is\ constant,\ y=y(x))\eqno(2)$$
are especially important for various applications.
\par
It is well-known that the Painlev\'e
transcendents are meromorphic functions which cannot be expressed in terms of
any known special functions (see \cite{6} ). Therefore particular attention
must be
paid to asymptotic (as $\vert x\vert\rightarrow\infty$) properties of the
Painlev\'e eqs.  solutions (see \cite{7,9,10} ). Authors of paper \cite{11}
developed an
asymptotic classification of solutions to the P1 eq.. Kapaev \cite{12} used
the isomonodromic deformations method (IDM) and considerably advanced in
this direction (while $x$ remains real). Moore, Novikov, Dubrovin and
Krichever \cite{13,14,15} applied some elements of finite-gap theory and
Whitham method to the problem mentioned above.

Studies in the theory of matrix model in two-dimensional gravity deal with
orthogonal polynomials with weight function

$$w(x)=exp[-V(x)],$$
where $V(x)$ is a polynomial in even powers. Equations

$$c^{1\over2}_{n}[V'(L)]_{n,n-1}=n \eqno(3)$$
where

$$L_{nm}=c^{1\over2}_{m}\delta_{n+1,m}+c^{1\over2}_{m}\delta_{n-1,m},\ n\in
{ \bf Z},$$
$\delta_{n,m}$ is the Kronecker symbol, are usually referred to as the
discrete string equations (see \cite{16} ). Problem of description for
asymptotic (as $n\rightarrow\infty$) behavior of solutions $c_{n}$ of (3)
was the subject of paper \cite{17} . The simplest nontrivial string
equation, first obtained in \cite{18} , corresponds to a case
$V(x)=g_{2}x^{2}+ g_{4}x^{4}$. Asymptotic (as $n\rightarrow\infty$)
formulae for $c_{n}$ while $g_{2}=0$ were written out by Nevai \cite{19} .
The main subject of this paper is to investigate the general ($g_{2}\neq
0$) case, when the equation (3) takes the following form:

$$c_{n}+4gc_{n}(c_{n-1}+c_{n}+c_{n+1})=\epsilon n+\nu;\
n=0,1,2,...,\eqno(4)$$
where $g,\ \nu,\ \epsilon$ are constant real-valued parameters, $c_{n}$ is
an unknown sequence. There is also a procedure (see \cite{16} ) of
transition to continuum limit, (so-called double-scaling limit)
transforming (4) into the Painlev\'e-1 equation (1):  $c_{n}=\rho
[1-2ay(x)],\ where\ \epsilon^{-1} =Ba^{5/2},\ \epsilon n=A(1+\delta a^2x),\
B=-72g,\ A=\rho /2,\ \delta =2/3,\ \rho =(24g)^{-1},\ under\ a\rightarrow
0$. Therefore the eq. (4) is usually referred to as Discrete Painlev\'e-1
Equation (DP1). Authors of \cite{20} investigated the double-scaling limit
from the IDM viewpoint (see detailed description of the method in \cite{6}
). Thus, the subject of the paper may be called "single-scaling limit" for
solutions to (4). Asymptotic (as $n\rightarrow\infty$) classification of
solutions of DP1 was developed recently in \cite{21} .  The classification
consists of the following.

\par {\bf Theorem 1.} The set of all solutions of the equation (4) consists
of the following three parts:

\par {\bf A.} There are two one-dimensional families of regular$\footnote[1]
{The terms "regular" and "singular" imply, roughly speaking, solutions
described by regular or singular asymptotic formulae at infinity,
correspondingly. Rigorous sense of these terms has been elucidated in the
paper \cite{21} }$ solutions, monotonous starting from some number, with
asymptotics $\pm\sqrt{{{n\epsilon}\over{12g}}}\ \footnote[2]{Suppose $g>$0.
Case $g<$0 leads to more tedious but principally analogous results}$

\par {\bf B.} Two-dimensional set of solutions singular at infinity.

\par {\bf C.} Countable set of one-dimensional families of "finite"
solutions \{$c_{0},\ c_{1},...,\ c_{N},\ 0$\}

It may be worthy of mentioning that classification of solutions to P1 (1),
(as $x \rightarrow -\infty$) developed in \cite{11} indicates a rather
similar picture:

1) 2-dimensional set of solutions singular at minus infinity;

2) 2-dimensional set of solutions with regular asymptotics:
$$
y_{-}(x) \sim -\sqrt{-{x \over 6}}\ as\ x \rightarrow -\infty
$$

3) 1-dimensional set of solutions $y_{+}(x) \sim -\sqrt{-{x \over 6}}\ as\
x \rightarrow -\infty$

The latter set of solutions (called "Physical Solutions") is of special
interest for applications. Asymptotic studies of solutions on real axis
from the IDM point of view for P1 were carried out by Kapaev \cite{12}, the
Whitham approach to this problem was developed by Krichever, Moore,
Dubrovin and Novikov \cite{13,14,15}

\par The main goal of this paper is to calculate asymptotics of real
solutions of the eq.(4) and to construct an approach to the problem from
the finite-gap theory and the Whitham method viewpoint. Asymptotic problem
for A-type solutions will be investigated by means of the IDM, and B-type
solutions will be studied in the last paragraph in the Whitham way.
However, one should note that both methods are powerful enough to
successfully treat solutions of the two types.

\par \vskip.5cm {\bf Linear problem}\vskip.5cm

\par Authors of papers \cite{20} obtained the Lax representation for (3):

$$ [L,A]=\epsilon,\hskip.2cm\epsilon>0,$$
where $L,A$ are linear operators in a sequence space.  One can easily verify
that the linear operators $L, A$ corresponding to eq.(4) have the form

$$L_{nm}=c^{1\over2}_{n}\delta_{nm-1}+c^{1\over2}_{m}\delta_{m+1n}\eqno(5)$$

$$A=2g(A^{1}+A^{2})+{1\over2}A^{3},\ where$$

$$A_{nm}^{1}=c^{1\over2}_{n}(c_{n-1}+c_{n}+c_{n+1})\delta_{n+1m}-c_{m}^
{1\over2}(c_{m-1}+c_{m}+c_{m+1})\delta_{m+1n},\eqno(6)$$

$$A^{2}_{nm}=(c_{n}c_{n+1}c_{n+2})^{1\over2}\delta_{n+3m}-
(c_{m}c_{m+1}c_{m+2})^{1\over2}\delta_{m+3n},$$

$$A^{3}_{nm}=c_{n}^{1\over2}\delta_{n+1m}-c_{m}^{1\over2}\delta_{m+1n}.$$
Thus, eq. (4) induces the appropriate couple of linear problems:

$$L\psi_{n}=\xi\psi_{n},\hskip.5cm \epsilon \partial_{\xi}\psi_{n}=A\psi_{n},
\eqno(7)$$
where $\xi$ is a spectral parameter, $\partial_{\xi}=\partial/\partial{\xi},\
\psi_{n}=\psi_{n}(\xi).$ Change the coordinates $\psi_{n}$ to $\Psi_{n}=
(c_{0}c_{1}...c_{n})^{1/2}\psi_{n}$ and rewrite the system (7) in matrix form:

$$\epsilon \partial_{\xi}\vec \Psi_{n}=V_{n}\vec \Psi_{n}\eqno(8a)$$

$$\vec \Psi_{n+1}={\cal L}_{n}\vec \Psi_{n},\ where\eqno(8b)$$

$$\vec\Psi_{n}=(\Psi_{n},-\Psi_{n-1})^{T},\hskip.5cm {\cal L}_{n}=
\pmatrix{\xi&c_{n}\cr-1&0},$$

$$V_{n}=2g\pmatrix{-\xi(\xi^2+2c_n+a)&-2c_n(\xi^2+c_n+c_{n+1}+a)\cr
2(\xi^2+c_{n-1}+c_n+a)&\xi(\xi^2+2c_n+a)},\eqno(9)$$ where
a=${1\over{4g}}$, exponent T indicates transposition.  The compatibility
condition for equations (8) looks so:

$$\epsilon\partial_{\xi}{\cal L}_n+{\cal L}_n V_n-V_{n+1}{\cal
L}_n=0,\eqno(10)$$
or, equivalently, eq. (4).

\par {\bf Remark.} The DP1 (4) describes self-similar solutions of
the Volterra lattice (29) and the second nontrivial member of the Volterra
lattice hierarchy (see [22]):

$$\partial_tf_n=f_nf_{n+1}(f_n+f_{n+1}+f_{n+2})-f_nf_{n-1}(f_{n-1}+
f_n+f_{n+1})$$

$$+af_n(f_{n-1}-f_{n+1})\eqno(11)$$
\par The purpose of the next paragraph is to calculate more precise
asymptotics for the regular solutions (A-type) of eq. (4).

\par\vskip.5cm {\bf Asymptotics of the regular solutions}\vskip.5cm

\par The formulae (8) indicate that eq. (4) can be treated as an equation of
isomonodromic deformations for the linear system of ODE with polynomial
coefficients

$$\partial_{\xi}\chi_n=V_n\chi_n,\eqno(12)$$
where $\chi_{n}$ is a 2$\ast $2-matrix. This allows to use the isomonodromic
deformations method described in [6]. The method consists of explicit
asymptotic calculation of the monodromy data of eq. (12). As follows from
\cite{20} , these monodromy data are invariant under dynamics of eq. (4).
Inversion
of the explicit formulae yields the asymptotics of solutions to eq. (4).

The eq. (12) has an irregular singular point at $\xi=\infty$. In an
neighborhood of infinity one can define eight canonical sectors
$\Omega_{k}$, k=1,2,...,8:

$$\Omega_k=\left\{\xi\in {\Bbb C}:{{2k-3}\over 8}\pi<arg\xi<{{k+1}\over
8}\pi \right\}$$
Canonical solutions of eq. (12), being determined in the appropriate
sectors $\Omega_k$ by their asymptotics

$$
\chi_{n}^{k}(\xi)\sim [I+O(\xi^{-1})]exp\left(\sigma_{3}\left({g \over
2}\xi^4 +{1 \over 4} \xi^2-nlog\xi\right)\right),\ \ \xi\in\Omega_k
$$
where $\sigma_{3}=\pmatrix{1&0\cr 0&-1}$, differ each from other only with
matrix multiplicators:

$$\chi_n^{k+1}(\xi)=\chi_n^{k}(\xi)S_k,\ where$$
$S_k $ is called Stokes matrix. Proof of the following theorem can be
extracted from paper \cite{20} :

\par {\bf Theorem 2.} Function $c_{n}$ satisfies the eq. (4) iff the
appropriate Stokes matrices do not depend on $n$.
\par
The Stokes matrices have the specific triangular structure:

$$S_{2k-1}=\pmatrix{1&s_{2k-1} \cr 0& 1},\ \ S_{2k}=\pmatrix{1&0\cr s_{2k}
&1}, \ \ k=1,2,3,4$$
Symmetry in the eq. (13) leads to constraint

$$\sigma_3 \chi_n^{k+4}(-\xi)\sigma_3=\chi_n^{k}(\xi),\ whence\ S_{k+4}=
\sigma_3 S_k \sigma_3,$$
which means together with the cyclic identity

$$S_{1}S_{2}...S_{8}=I$$
that there are only two independent elements of the Stokes matrices ($s_2$ and
$s_3$).
\par Denote $\chi_{n}^{k}=T_{n}\Phi_{n}^{k}$, where $T_{n}$ reduces $V_{n}$
to a diagonal form, and get:

$$\partial_{\xi}\Phi_{n}^{k}=\left(\Lambda_n-T_{n}^{-1}\partial_{\xi}
T_n\right)\Phi_{n}^{k}, \eqno(13)$$
where $\Lambda_n=\mu (\xi)\sigma_{3};\hskip.3cm \mu (\xi)$ is an eigenvalue
of the matrix $V_{n}                                   .$
$T_n=\pmatrix{1 & t_1 \cr t_2 & 1}$, where $t_1=(-v_{11}+\mu)/v_{21},\
t_2=(v_{11}-\mu)/v_{12},\ v_{ij}$ are entries of matrix $V_n$. One easily
evaluates: $t_1 \sim c_n \xi^{-1}+O(\xi^{-3}),\ t_2 \sim \xi^{-1}
+O(\xi^{-3})$ as $\xi \rightarrow \infty$, whence
$$
\Phi_{n}^{k}(\xi)\sim [I+O(\xi^{-1})]exp\left(\sigma_{3}\left({g \over
2}\xi^4 +{1 \over 4} \xi^2-nlog\xi\right)\right),\ \ \xi\in\Omega_k
$$
and
$$\Phi_{n}^{k+1}(\xi)=\Phi_{n}^{k}(\xi)S_k,$$

\par Let us start now the asymptotic (as n$\rightarrow\infty$)
investigation of the direct monodromy problem in case of the regular
solution of eq. (4) with asymptotics $c_{n}\sim
\sqrt{{n\epsilon\over{12g}}}.$

\par Introduce in eq. (12) notation

$$\tau={{an}\over3},\ \xi=\tau^{1/4}\lambda,\ c_n=\tau^{1/2}(1+u_n),\
where\hskip.2cm a=(4g)^{-1},\ u_n=O(\tau^{-1/2})$$
and get:

$${\partial_{\lambda}}{\chi_n}=\tau{A_n}{\chi_n},\eqno(14)$$ where

$$A_n=\left(\matrix{-\lambda(\lambda^2+2+u_{n-1}+u_n+a\tau^{-1/2})\cr
2\tau^{-1/4}(\lambda^2+2+u_{n-1}+u_n+a\tau^{-1/2})}$$

$$\left{\matrix{-2\tau^{1/4}(1+u_n)(\lambda^2+2+u_{n+1}+u_n+a\tau^{-1/2})
\cr \lambda(\lambda^2+2+u_{n-1}+u_n+a\tau^{-1/2})}\right)$$

$$\partial_\lambda\Phi_n(\lambda)=\tau\left(\Lambda_n-\tau^{-1}T_n^{-1}
{\partial_\lambda}{T_n}\right)\Phi_n(\lambda),\eqno(15)$$ where
$\Lambda_{n}=\tilde\mu(\lambda)\sigma_{3}$, matrix $T_{n}$ reduces $A_{n}$
to the diagonal form. $\tilde\mu(\lambda)$ is an eigenvalue of the matrix
$A_{n}$:

$$(2a\tilde\mu)^2=(\lambda^2+2)^2(\lambda^2-4)+2a\tau^{-1/2}(\lambda^4-4)+
\lambda^2a^2\tau^{-1}-$$

$${2\over3}a^2\tau^{-1}+12u^2_n(\lambda^2+2)-c^2+O(\tau^{-3/2}\lambda^0),
\eqno(16)$$

$$-c^2=2\left[(1+\sqrt3)u_n+(\sqrt3-1)u_{n+1}+{1\over{\sqrt3}}a\tau^{-1/2}
\right]\times$$

$$\left[(1+\sqrt3)u_{n+1}+(\sqrt3-1)u_n+{1\over{\sqrt3}}a\tau^{-1/2}
\right],$$

$$u_{n-1}+4u_n+u_{n+1}+a\tau^{-1/2}=O(\tau^{-1})$$
Turning points of the system (15), i.e. zeros of function (16), can be found
in the following way:

$$\lambda_{1,2,3,4}=\pm
i\sqrt2+O(\tau^{-1/2}),\hskip.5cm\lambda_{5,6}=\pm2+
O(\tau^{-1/2})\eqno(17)$$
The points $\lambda_{1,2}\ and\ \lambda_{3,4}$ coincide in the limit
$\tau\rightarrow\infty$, the points $\lambda_{5,6}$ are simple.\par One can
compute asymptotics of solutions of eq. (15) as $\lambda\rightarrow\infty$:

$$\Phi^k_n(\lambda)\sim [I+O(\lambda^{-1})]\times\eqno(18)$$

$$exp\left\{{{\tau\sigma_3}\over{2a}}\left[{1\over4}\lambda(\lambda^2+2)
\sqrt{\lambda^2-4}}-6log\left({\lambda+\sqrt{{\lambda^2-4}}\over2}\right)+
{a\over 2}\tau^{-1/2}\lambda\sqrt{\lambda^2-4}\right]\right\},$$

$$\lambda\in\Omega_k$$
The procedure for the monodromy data calculation is as follows. First look for
approximate WKB-solutions of the problem (15):

$$\Phi^k_{WKB}(\lambda)=exp\left\{\tau\int\limits_{\lambda_0}
^{\lambda}\left(\Lambda_n-\tau^{-1}T_n^{-1}\partial{_\lambda}T_n\right)
d\lambda\right\}\eqno(19)$$ Near the turning points $\lambda_{1,2}$ define
a region

$$D_k=\left\{\lambda-i\sqrt2=\zeta=O\left(\tau^{\delta-1/2}\right)\right\}
\cap \Omega_k,\ where\ \delta>0.$$
In this region the WKB-solution (19) satisfies the eq.(15) up to a quantity of
order $O(\tau^{-2\delta})$. Connect the solution (18) with the WKB-solution
(19) in the following manner:

$$\Phi^k_n(\lambda)=\Phi^k_{WKB}(\lambda)C_k(\lambda)\eqno(20)$$
In the same neighborhood of the turning points $\lambda_{1,2}$ one can obtain
another approximate solution in terms of the parabolic cylinder functions (see
[23]). So, eq. (15) in the region $D_{k}$ gets the form

$$\partial_z\tilde\Phi=\pmatrix{z/2&\beta\cr \gamma
&-z/2}\tilde\Phi,\eqno(21)$$
where $z=3^{1/4}2(\zeta+\eta)(\tau/a)^{1/2}$,
$\eta={{ia}\over{3\sqrt{2\tau}}}$

$$\beta={1\over2}\sqrt[4]{\tau\over3}\sqrt{\tau\over{2a}}\left[
(1+\sqrt3)u_n+(\sqrt3-1)u_{n+1}+{1\over{\sqrt3}}a\tau^{-1/2}\right]$$

$$\gamma={1\over2}\sqrt[4]{1\over{3\tau}}\sqrt{\tau\over{2a}}\left[
(\sqrt3-1)u_n+(\sqrt3+1)u_{n+1}+{1\over{\sqrt3}}a\tau^{-1/2}\right]$$
System (21) can be explicitly solved in terms of Weber-Hermit functions. In the
region $D_{k}$ this solution connects with the WKB-solution:

$$\Phi^k_{WKB}(\lambda)=\tilde\Phi(\lambda)N_k(\lambda)\eqno(22)$$
and, therefore, $\Phi^k_n(\lambda)=\tilde\Phi^k_n(\lambda)N_k(\lambda)
C_k(\lambda)$. Thus,

$$S_2S_3=\left[\Phi^2_n(\lambda)\right]^{-1}\Phi^4_n(\lambda)=
C_2^{-1}(\lambda)N_2^{-1}(\lambda)N_4(\lambda)C_4(\lambda)\eqno(23)$$
One can easily calculate matrix $N_2^{-1}(\lambda)N_4(\lambda)$ using the
known asymptotics of the Weber-Hermit functions as
$\vert{z}\vert\rightarrow\infty$ (see \cite{23} and cf. \cite{12} ):

$$N_2^{-1}(\lambda)N_4(\lambda)=\pmatrix{1&-\beta\sqrt{2\pi}exp(-\pi
i\rho)[\Gamma(1+\rho)]^{-1}\cr -i\sqrt{2\pi}[\beta\Gamma(-\rho)]^{-1}&
exp(-2i\pi\rho)},\eqno(24)$$
where $\rho=\beta\gamma=-{{\tau c^2}\over{16\sqrt3a}}$ (see (16)), $\Gamma$
is Euler $\Gamma$-function.

\par Now determine the matrix $C_{k}$ (see (20)). First compute an integral in
formula (19), where $\lambda_{0}=i\sqrt{2}-\eta$ :
$\int\limits_{i\sqrt2-\eta}\limits^\lambda\mu(\lambda)d\lambda=I_1+I_2,$ where

$$I_1=\int\limits_{i\sqrt2-\eta}\limits^{i\sqrt2+i\Delta}
\mu(\lambda)d\lambda,\hskip.5cm
I_2=\int\limits_{i\sqrt2+i\Delta}\limits^\lambda
\mu(\lambda)d\lambda,\hskip.5cm \Delta=\tau^{-1/2}\left(1-{a\over{3\sqrt2}}
\right);$$

$$2aI_1={1\over{4\sqrt3}}\left[{1\over2}y^2-{c^2\over2} log
y\right]_{y=4\sqrt3i\tau^{\delta-1/2}}
+{c^2\over{8\sqrt3}}\left(log\sqrt{{{-c^2}\over2}}-{1\over2}\right),$$

$$2aI_2=$$

$$\left[{1\over4}\lambda(\lambda^2+2)\sqrt{\lambda^2-4}-6log\left({\lambda+
\sqrt{{\lambda^2-4}}\over2}\right)+{a\tau^{{-1/2}}\over2}\lambda
\sqrt{\lambda^2-4} +O(\tau^{-3/2}\lambda^{-1}){\right
]}^\lambda_{i\sqrt2+i\Delta},$$
whence obtain:

$$C_k=exp\left\{{{{\sigma_3\tau}\over{2a}}H\right\},\ where$$

$$H=3i\pi+log(2+\sqrt3)+\sqrt3a\tau^{-1/2}-2\sqrt3\Delta^2+
2\sqrt{{2\over3}}\Delta a\tau^{-1/2}+{{2\sqrt{3}}\over{\tau}}+$$

$${{a\rho}\over{\tau}}\left(log\rho+log a-{1\over2}
log3-2log2-1-i\pi\right)$$ and then, using (23,24):

$$S_2S_3=\pmatrix{1&s_3\cr s_2&1+s_2s_3}=\pmatrix{1&-{{\beta\sqrt{2\pi} exp
(f-i\pi\rho)}\over{\Gamma(1+\rho)}}\cr -{{i\sqrt{2\pi} exp(-f)}
\over{\beta\Gamma(-\rho)}}&exp(-2i\pi\rho)},\eqno(25)$$

$$f=-n[i\pi+log(2+\sqrt3)]-\sqrt{3\tau}+{{2\tau}\over a}\Delta\left(
\sqrt3\Delta-\sqrt{{{2}\over{3}}}a\tau^{-1/2}\right)-{{2\sqrt3}\over a}$$

$$+{1\over4}log\tau+\rho\left({1\over2}log3+2log2-log a\right);$$

$$\rho=\beta\gamma,\hskip.5cm \Delta=\tau^{-1/2}\left[1-
{a\over{\sqrt{18}}}\right]$$

$$\beta=-{{\sqrt{2\pi}}\over{s_2\Gamma(-\rho)}}exp\left({{i\pi}\over2}-
f\right),\hskip.5cm
\gamma=-{{\sqrt{2\pi}}\over{s_3\Gamma(\rho)}}exp(-i\pi\rho+f)$$
Utilizing (21), one gets:

$$u_n\sim-{{\sqrt3+1}\over{\sqrt3}}{{\sqrt{\pi}}\over{s_2\Gamma(-\rho)}}
\sqrt{{a\over\tau}}3^{1/4}$$

$$
\times exp\left({{i\pi}\over2}-f\right)+
{{\sqrt3-1}\over{\sqrt3}}{{\sqrt{\pi}}\over{s_3\Gamma(\rho)}}
\sqrt{{a\over\tau}}3^{1/4}exp(-i\pi\rho+f)-{{a\tau^{-1/2}}\over6}+O(\tau^{-1})
$$

Therefore,

$$u_n\sim-{{a\tau^{-1/2}}\over6}+\sqrt{{a\over{3\tau}}}[d_1 exp(f) +d_2
exp(-f)],\eqno(26)$$
where $d_1=3^{1/4}{{\sqrt3-1}\over{\sqrt3}}{{\sqrt{\pi}}\over{s_3
\Gamma(\rho)}},\hskip.5cm d_2=-3^{1/4}{{\sqrt3+1}\over{\sqrt3}}{{\sqrt{\pi}}
\over{s_2\Gamma(-\rho)}}exp(-i\pi\rho+{{i\pi}\over2}),$\par\vskip.3cm
$d_1d_2=-\sqrt3\rho;\hskip.5cm \rho={i\over{2\pi}}log(1+s_2s_3).$

\par The condition $u_{n}=O(\tau^{-1/2})$ corresponds to a particular case
$s_{3}=0\ :$

$$d_{1}=3^{1/4}\frac{\sqrt3-1}{2\sqrt{\pi}}s_{2}i,\hskip.5cm d_{2}=0.$$
The final formula gets the following form

$$c_n\sim\tau^{1/2}-{a\over6}+\sqrt{{a\over3}}3^{1/4}{{\sqrt3-1}\over
{2\sqrt{\pi}}}\times$$

$$i s_2 exp\left\{-n\left[\pi i+log(2+\sqrt3)\right]-\sqrt{3\tau}
+{{2\tau\Delta}\over{a}}\left[\Delta\sqrt3-{{a}\over{\sqrt{3\tau/2}}}\right]
-{{2\sqrt3}\over{a}} +{1\over4}log\tau\right\}\eqno(27)$$

$$
+O(\tau ^{-1/2})
$$
The exponentially small terms are presented in (27) to indicate the role of
the free parameter $s_2$ in the asymptotic series.

\par Now calculate asymptotics of solutions of eq. (4) with leading term
$c_{n}\sim -\tau^{1/2}$. First find eigenvalues of the matrix $A_n$ (14),
where $c_{n}=\tau^{1/2}(-1+u_{n}),\hskip.5cm u_{n}=O(\tau^{-1/2})\ :$

$$(2a\mu)^2=(\lambda^2-2)^2(\lambda^2+4)+2a\tau^{-1/2}(\lambda^4-4)+
\lambda^2a^2\tau^{-1}+{2\over3}a^2\tau^{-1}$$

$$+12u^2_n(\lambda^2-2)-c^2+4b_n(\lambda^2-2)+O(\tau^{-3/2}\lambda^0),\
where$$

$$b_n=u_{n-1}+4u_n+u_{n+1}+a\tau^{-1/2}=O(\tau^{-1})$$

$$c^2=2\left[(1+\sqrt3)u_n+(\sqrt3-1)u_{n+1}+{1\over{\sqrt3}}a\tau^{-1/2}
\right]$$

$$\times
\left[(1+\sqrt3)u_{n+1}+(\sqrt3-1)u_n+{1\over{\sqrt3}}a\tau^{-1/2}\right]$$
The remaining calculations are analogous to those accomplished above for
the solutions with leading term $c_{n}\sim\tau^{1/2}$, where the connection
region must be chosen near the turning  points
$\lambda_{1,2}=\sqrt2+O(\tau^{-1/2})$.

$$C_k=exp\left\{{{{\sigma_3\tau}\over{2a}}H\right\},\ where$$

$$H=3i\pi-3 log(2+\sqrt3)-
\sqrt3a\tau^{-1/2}-2\sqrt3\Delta^2-2\sqrt{{2\over3}}\Delta a\tau^{-1/2}$$

$$-2\sqrt3\tau^{-1}+a\rho\tau^{-1}\left(log\rho+log-1-{1\over2}log3
-2log2-1\right)$$
Then one obtains:

$$c_n\sim-\tau^{-1/2}}-{a\over6}+\sqrt{{a\over3}}[d_1 exp(f) +d_2
exp(-f)],$$
where

$$d_1=3^{1/4}(\sqrt3-1){{\sqrt{\pi}}\over{s_3\Gamma(\rho)}}
 exp(-i\pi\rho),\hskip.5cm
d_2=-3^{1/4}(\sqrt3+1)i{{\sqrt{\pi}}\over{s_2\Gamma(-\rho)}},$$

$$\Delta=\tau^{-1/2}\left(1-{a\over{3\sqrt2}}\right);$$

$$f=n[i\pi+log(2+\sqrt3)]+\sqrt{3\tau}+{{2\tau}\over a}\Delta\left(
\sqrt3\Delta+\sqrt{{{2}\over{3}}}a\tau^{-1/2}\right)+{{2\sqrt3}\over a}$$

$$+{1\over4}log\tau+\rho\left({1\over2}log3+2log2-log\right).$$
The final formula for asymptotics has the following form:

$$c_n\sim-\tau^{1/2}-{a\over6}-\sqrt{a\over3}3^{1/4}{{\sqrt3+1}
\over{2\sqrt{\pi}}} s_3\times$$

$$ exp\left\{-n\left[i\pi +log(2+\sqrt3)\right]-\sqrt{3\tau}
-{{2\tau\Delta}\over{a}}\left[\Delta\sqrt3+{a\over{\sqrt{3\tau/2}}}\right]-
{{2\sqrt3}\over{a}}-{1\over4}log\tau\right\},\eqno(28)$$
$$
+O(\tau ^{-1/2})
$$
where $\tau={{an}\over3},\ \Delta=\tau^{-1/2}
\left(1-{a\over{3\sqrt2}}\right),\ s_2=0. $
The exponentially small terms are presented in (28) to indicate the role of
the free parameter $s_3$ in the asymptotic series.
The regular solutions of (4) can be treated via the quasiclassical
(Whitham) point of view as well (see Theorem 5 lower).
\vskip.3cm
{\bf Proposition.}

(i) There exist two unique regular "algebraic" solutions to (4):
$c_{n}^{\pm} \sim \pm \sqrt{{n}\over{12g}} +O(1)$ as $n\rightarrow \infty$,
decomposable at infinity into asymptotic series in purely power terms.
These terms can be computed via simple iteration procedure.

(ii) There exist two one-parametric families of regular (type A) solutions.
These families are characterized by asymptotic formulae (27) and (28),
where $s_2$ and $s_3$ are free parameters correspondingly. Any solution of
the two families exponentially quickly tends to appropriate algebraic
solution: $c_{n}^{+}$ or $c_{n}^{-}$.

\par Next paragraph concerns application of
the Whitham method to the problem of asymptotic integration of DP1 (4).

\par\vskip.5cm {\bf Quantization of one-gap potentials}\vskip.5cm

\par Note certain connection between the DP1 (4) and the Volterra lattice

$$\partial_{t}v_{n}=v_{n}(v_{n+1}-v_{n-1}),\hskip.2cm n\in{\Bbb Z},\
v_{n}=v_{n}(t),\ t\in{\Bbb R}\eqno(29)$$
Suppose the functions $\vec\Psi_{n},\ c_{n}$ (8) depend on time $t\in{\Bbb
R}\ : \vec\Psi_{n}=\vec\Psi_{n}(\xi,\ t),\ c_{n}=c_{n}(t)$. Determine their
t-dynamics:

$$\partial_{t}\vec\Psi_{n}=W_{n}\vec\Psi_{n},\eqno(30)$$
where operator $W_{n}$ has been written out in \cite{24} :

$$W_{n}=\left(\matrix{-{1\over2}\xi^4-\xi^2(c_{n}+{{a}\over2})-c_{n}(c_{n-1}+
c_{n}+c_{n+1}+a)& \cr \xi(\xi^2+c_{n}+c_{n-1}+a)}\right}$$

$$\left{\matrix{-{\xi}c_{n}(\xi^2+c_{n}+c_{n+1}+a)\cr
{1\over2}\xi^4+\xi^2(c_{n}+{{a}\over2})-c_{n-1}(c_{n-2}+c_{n-1}+c_{n}+a)}
\right) \eqno(31)}$$ Putting $\epsilon=0$ in eq. (10) and using (8,30), one
gets the standard one-gap integration procedure (see \cite{24} ) which yields
exact
solutions of the eq. (4) while $\epsilon=0$ :

$$c_{n}^0=\zeta(nk+\omega')-\zeta(nk-k+\omega')+\zeta(2k)-\zeta(k)\
\footnote[3]{see about Weierstrass elliptic functions in \cite{25} }
\eqno(32)$$

$$
\zeta '(x)=-\wp (x),\ x=\int\limits^{\wp(x)}\limits_{\infty}
{{dx}\over{\sqrt{4x^3 -g_2 x-g_3}}},\ \zeta (x)={{\sigma '(x)}\over{\sigma
(x)}} $$ $2\omega,\ 2\omega'$ are periods of function $\wp (x),\
\eta=\zeta(\omega),\ \eta'=\zeta(\omega')$; $g_2,\ g_3,\ k$ are parameters.
The formula (32) corresponds to the one-gap solution of the Volterra lattice
(29). Three parameters that completely determine the function (32) can be
interpreted as three nontrivial branch points of genus-one Riemann surface
$\Gamma^{0}(w,\lambda)$ :

$$w^2=R_{4}(\lambda)=-detW_{n}=\lambda(\lambda^3+2a\lambda^2-4\lambda
C_{n}^0-4D_{n}^0),\eqno(33)$$

$$where\ \lambda=\xi^2,\hskip.5cm C^0_{n}={\nu}a-{{a^2}\over4},\hskip.5cm
D^0_{n}=c_{n}^0(c_{n-1}^0+c_{n}^0+a)(c_{n+1}^0+c_{n}^0+a)\eqno(34)$$
Then put $\epsilon$ a small positive-valued number and in force of (4) obtain
a Riemann surface $\Gamma^{\epsilon}(w,\lambda)$ :

$$w^2=R_{4}^{\epsilon}(\lambda)=\lambda(\lambda^3+2a\lambda^2-4\lambda
C_{n}^{\epsilon}-4D_{n}^{\epsilon}),\eqno(35)$$
whose branch points "slowly" depend on n because of the coefficients dynamics

$$C_{n}^{\epsilon}=C_{n}^0+{\epsilon}n,\hskip.5cm
\Delta_{n}D_{n}^{\epsilon}=
D_{n+1}^{\epsilon}-D_{n}^{\epsilon}=\epsilon(c_{n+1}+c_{n}+a)\eqno(36)$$
Naturally, the perturbed Riemann curve (35) specifies potential that differs
from the initial one $c_n$ (32).

\par The Whitham method investigates solutions, determined by parameters
depending on slow continuous variables, so introduce the following
pseudodifferential continuum equation

$$c(x)+4gc(x)[c(x-1)+c(x)+c(x+1)]={\epsilon}x,\hskip.3cm x\in{\Bbb
R}\eqno(37)$$
where $c(x)$ is a piecewise smooth function, c(n)=$c_n,\ n\in{\Bbb Z}$.  We
shall examine function (32), where $x\in{\Bbb R}$ is substituted instead of
$n\in{\Bbb Z}$ and the three parameters determining (32) depend on "slow"
variable X=x$\epsilon$ in force of eqs. (36). The problem now is to describe
appropriate dynamics of the branch points.

\par A system describing slow drift of the branch points in variable $X$ is
usually referred to as modulation equations or Whitham system.

{\bf Theorem 3.} The modulation equations can be written out in
"Flashka-Forest-McLaughlin-Krichever form" (cf. \cite{26,27} )

$$\partial_{X}w=-2a\lambda\partial_{\lambda}p, \footnote[4]{similar type of
modulation eqs. for P1 was obtained in [14]} \eqno(38)$$

$$where\hskip.5cm X={\epsilon}x,\hskip.5cm
w^2=\lambda(\lambda^3+2a\lambda^2+{\lambda}G_{2}+G_{3}),$$

$$G_{2}=a^2-4\nu-4X,\hskip.5cm G_{3}=-4D,\hskip.5cm
\partial_{X}D=2\bar{c}+a, \hskip.5cm a={1\over{4g}}$$

$$g\ is\ a\ constant,\ \lambda \ is\ a\ spectral\ parameter;$$

$$p(z)={{i\pi}\over{\omega\omega'}}\left[log\frac{\sigma(z+k)}
{\sigma(z-k)}-2kz\eta'/\omega'\right]\eqno(39)$$
is a quasi-impulse,

$$\lambda(z)=\zeta(z-k)-\zeta(z+k)+2\zeta(k),\hskip.5cm
w(z)=\wp(z+k)-\wp(z-k);$$ $\bar c$
is a mean of $c$.

Quantity $\oint\limits_{b}{{w}\over{\lambda}}d\lambda$ is an integral of the
eqs. (38), $b$ is a cycle over gap on $\Gamma^{\pm}$.

Proof. Using the representation (10) one can compute dynamics in $n$ of
eigenvalues $E_{n}$ of the matrix $V_{n}$ :

$$\Delta_{n}E_{n}=E_{n+1}-E_{n}=2a\epsilon\frac{\langle
E_{n+1}\vert\partial _{\xi}{\cal L}_{n}\vert{E_{n}}\rangle}{\langle
E_{n+1}\vert{\cal L}_{n}\vert {E_{n}}\rangle}=
-2a\epsilon\frac{\lambda+c_{n+1}+c_{n}+a}{E_{n}}+O(\epsilon^2)$$
Average and obtain: $\partial_{X}E=-2a\frac{\lambda+2\bar c+a}{E}$. Then,
taking into account that $E=w/\sqrt\lambda$ , one gets the claim. Identity
$\partial_{X}D=2\bar c+a$ ensues from (36).

$$\partial_{X}\oint\limits_{b}{w\over{\lambda}}d\lambda=
-2a\oint\limits_{b}dp(\lambda)=0$$

The following ideology for Theorem 4 is quite similar to that developed by
Krichever for P1 (see [28]).

The purpose of the Whitham method here is to look for solution of (37) in
the form

$$
c(x)=c^{0}_{x}(\epsilon ^{-1} S(X)+\Phi (X);\ \bar{\lambda}(X))+O(\epsilon),
$$
where elliptic function $c^{0}_{x}$ is specified by formula (32),
$\bar{\lambda}=(\lambda_1,\ \lambda_2,\ \lambda_3,\ \lambda_4)$ are branch
points of (35), $\partial_X S=k;\ \Phi$ is a phase shift. $\bar{\lambda}(X)$
drifts in force of the Whitham system (35-36), (38) or (40). At the same
time one can choose variable $x$ large enough so that $\epsilon x=X
\rightarrow\infty$. Therefore there exists a rescaling procedure $x
\rightarrow \epsilon^{-1-\alpha} x,\ \alpha >0$ which provides the following
fact: the function $c^{0}_{x}$ (32), where $n$ is replaced by $x$ and
parameters drift in force of (38), represents a leading term of asymptotic
series (in $x^{-1} \rightarrow 0$) to solution of equation (37) for
$\epsilon =O(1)$. Thus, we have

{\bf Theorem 4.} The formally elliptic function (32), whose parameters are
specified by system (38), is a leading term of asymptotic series to solution
of (4) for $n \rightarrow \infty$ and $\epsilon=O(1)$.

Now we develop a kind of classification for solutions of (4) in terms of the
appropriate branch points (see (32-36)), i.e. give an answer to the question:
which solutions of the modulation eqs. (38) correspond to the solutions of eq.
(4) of A - and B types?

\par {\bf Theorem 5.} Let's pose the Cauchy problem for eq. (4) started from
some large number $n_{0}$. Its solutions with asymptotics
$c_{n}^{\pm}\sim\pm\sqrt{\frac{na\epsilon}{3}}$ correspond to real Riemann
surfaces of genus one $\Gamma^{\pm}(w,\ \lambda):\ w^2=R^{\pm}_{4}(\lambda,\
\epsilon n_{0},\ D)=\lambda R^{\pm}_{3}(\lambda,\ \epsilon n_{0},\ D)$, where
the polynomials of the third order $R^{\pm}_{3}$ are specified by conditions

(i) $R^{\pm}_{3}(\lambda,\ \epsilon n_{0},\ D)$ has one zero near point
$\lambda^{\pm}_{0}=\pm4 u_{0}(n_{0}),\ where\  u_{0}(n)=
\sqrt{\frac{na\epsilon}{3}}$;

(ii) maximum (minimum) of R^{\pm}_{3}(\lambda,\ \epsilon n_{0},\ D)$  is
located near point $\lambda^{\pm}=\mp 2u_{0}(n_{0})$.

To prove the claim one needs only to substitute formulae (27,28) into (35) and
to examine arising series:

$$D^{\pm}=\pm4u_{0}^{3}+2au_{0}^{2}\pm2u_{0}({{a^2}\over{12}}+\nu)+...,$$

$$R^{\pm}_{3}(\lambda,{\epsilon}n_{0},D^{\pm})\vert_{\lambda\rightarrow
\mp2u_{O}}\sim\mp{8\over3}a^2u_{0}+...$$

{\bf Consequence.} Solutions of (4) with asymptotics
$\pm\sqrt{\frac{na\epsilon} {3}}$ (A-type) correspond to Riemann curves
$\Gamma^{\pm}$ of genus one with four branch points $\lambda^{\pm}_{1,2,3,4}$
:

$$Re(\lambda^{\pm}_{1,2})\sim-2u_{0}<\lambda^{+}_{3}=0<\lambda^{+}_{4}
\sim4u_{0},$$

$$\lambda^{-}_{1}\sim-4u_{0}<\lambda^{-}_{2}=0<Re(\lambda^{-}_{3,4})
\sim2u_{0},\ where$$

$$\lambda^{+}_{3,4},\ \lambda^{-}_{1,2}\in{\Bbb R};\hskip.5cm
\lambda^{-}_{3}=\bar{\lambda^{-}_{4}},\hskip.5cm \lambda^{+}_{1}=\bar
{\lambda^{+}_{2}}$$
bar means complex conjugation, $u_{0}=\sqrt{\frac{na\epsilon}{3}}$.

\par{\bf Remark.} Eq. (38) is equivalent to the following system of ODE on the
branch points:

$$\partial_{X}\lambda_{1}=4a\frac{\lambda_{1}\Pi}{(\lambda_{3}-\lambda_{1})
(\lambda_{4}-\lambda_{1})K}$$

$$\partial_{X}\lambda_{2}=4a\lambda_{2}\frac{K-\Pi}{(\lambda_{3}-\lambda_{2})
(\lambda_{4}-\lambda_{2})K}$$

$$\partial_{X}\lambda_{3}=4a\lambda_{3}\frac{K-(1-c^2)\Pi}
{(\lambda_{3}-\lambda_{2})(\lambda_{3}-\lambda_{4})K}\eqno(40)$$

$$\partial_{X}\lambda_{4}=4a\lambda_{4}\frac{K\tilde{k^2}-(\tilde{k^2}-c^2)
\Pi}{(\lambda_{4}-\lambda_{1})(\lambda_{4}-\lambda_{3})Kc^2},$$
where $\lambda_{j}$ are the branch points of $\Gamma(w,\lambda);\ c^{2}=
\frac{\lambda_{3}-\lambda_{2}}{\lambda_{3}-\lambda_{1}},\
\tilde{k}^{2}=c^{2}\frac{\lambda_{4}-\lambda_{1}}{\lambda_{4}-\lambda_{2}};\
K,\ \Pi $ are complete elliptic integrals:

$$K=\int\limits_0\limits^1\frac{dx}{\sqrt{G(x)}},\hskip.3cm
\Pi=\int\limits_0\limits^1\frac{dx}{(1-c^{2}x^{2})\sqrt{G(x)}},$$

$$G(x)=(1-x^2)(1-\tilde{k^2}x^2).$$
In fact, only one coefficient of the polynomial (35) has non-trivial Whitham
dynamics, so the system (40) can be represented by a single equation, for
example,

$$
\partial_X \bar{D} = 2\bar{c}+a,
$$
where $D=c_n(c_{n-1}+c_{n}+a)(c_{n}+c_{n+1}+a)$.

{\bf Lemma.} The system (40) has following two real-valued solutions
($\lambda_{1}<\lambda_{2}<\lambda_{3}<\lambda_{4})\ with\ asymptotics\
(X\rightarrow\infty)$:

$$(1).\ \lambda_{1}(X){\sim}-2\sqrt{aX},\ \lambda_{2}(X)
{\sim}d_{1}X^{-1/2},\ \lambda_{3}\equiv 0,\
\lambda_{4}(X){\sim}2\sqrt{aX},\eqno(41)$$

$$(2).\ \lambda_{1}(X){\sim}-2\sqrt{aX},\ \lambda_{3}
(X){\sim}d_{2}X^{-1/2},\ \lambda_{2}{\equiv} 0,\
\lambda_{4}(X){\sim}2\sqrt{aX},$$

Now we compute asymptotics for solutions of (4) that match solutions (41) of
the modulation eqs. (40).

{\bf Theorem 6.} Singular solutions of the eq. (4) (type B) correspond (see
the Theorem 4) to dynamics of the branch points $\lambda_{j}$,\ j=1,2,3,4,
specified by solutions (41) of the modulation equations (40). The appropriate
asymptotic formulae have form:

$$c_{2n}=-\sqrt{2a{\epsilon}n}\ tan \left[
{{n\pi}\over{2}}+\sqrt{a{\epsilon}n} +b+O(n^{-1/2})\right]+O(1)\eqno(42)$$

$$c_{2n+1}=-\sqrt{2a{\epsilon}n}\ cot\left[ {{n\pi}\over{2}}+
\sqrt{a{\epsilon}n} +b+O(n^{-1/2})\right]+O(1)$$
where $a={1\over{4g}}$, $b$ is a parameter.

Proof. Solution of eq. (4) is represented as

$$c_{n}=\zeta(z+k)-\zeta(z-k)+\zeta(2k)-\zeta(4k)\eqno(43)$$
where z=2nk, $a={1\over{4g}}=2\zeta(4k)-4\zeta(2k)$. The uniformization
($\lambda$(z), w(z)) transforming Riemann surface $\Gamma$ (35) to standard
torus $T^2$ gets the form (due to (43)):

$$\lambda(z)=\zeta(z+2k)-\zeta(z-2k))-2\zeta(2k)\eqno(44)$$

$$w(z)=\wp(z-2k)-\wp(z+2k)$$

Identity $c_{n}+c_{n+1}=\lambda(2nk)-a$ ensues from (43, 44). Using equality
$z=\int\limits_{\infty}\limits^{\lambda(z)}\frac{dt}{\sqrt{R_{4}(t)}}$, one
 can calculate asymptotics of function $\lambda(z)$ with the branch points
 determined by formulae (41). The leading term:

$$z\sim\int\limits_{\infty}\limits^{\lambda(z)}\frac{dt}{t\sqrt{(t-\lambda_{1})
(t-\lambda_{4})}}\sim{1\over{\sqrt{aX}}}\left[arctan( exp\phi(z))-
{{\pi}\over{2}}\right],$$
where $coth\phi(z)={{\lambda(z)}\over{2\sqrt{aX}}}.$ Then reduce:

$$\lambda(z)=-\sqrt{aX}[tan(z\sqrt{aX})+ cot(z\sqrt{aX})]+O(1)$$

To obtain asymptotics of the wave number $k(X)$ one should use
formula

$$k=-2\int\limits_{\lambda_{1}}\limits^{\lambda_{2}}p_{\lambda}d\lambda,
\eqno(45)$$
where $p$ is the quasi-impulse  (see (39)):

$$dp(\lambda)=\frac{\lambda+d}{\sqrt{R_{4}(\lambda)}}d\lambda,$$
$d$ is a constant determined by constraint $\int\limits_{\lambda_{2}}
\limits^{\lambda_{3}}dp=0$. Evaluation of the integral (45) yields asymptotics
 $k(X)=\frac{\pi}{8\sqrt{aX}}+O(X^{-1})$. Therefore,

$$c_{n}+c_{n+1}=-\sqrt{a{\epsilon}n}\left[tan\left(
{{n\pi}\over{4}}+\sqrt{\frac{
a{\epsilon}n}{2}}+...\right)+cot\left({{n\pi}\over{4}}+\sqrt{\frac{
a{\epsilon}n}{2}}+...\right)\right]+O(1)\eqno(46)$$
Conditions (35, 36) indicate that

$$D_{n}=c_{n}(c_{n-1}+c_{n}+a)(c_{n+1}+c_{n}+a)=(aX+\nu-c_{n}c_{n+1})
(c_{n+1}+c_{n}+a)$$
Now use (46):

$$c_{n}\sim{1\over2}\left\{\lambda(2n)-a\pm\left([\lambda(2n)-a]^{2}-
4\left[aX+\nu+\frac{D sin(4kn\sqrt{aX})}{2\sqrt{aX}}\right]\right)^{1/2}
\right\}$$
One obtains at the leading term:

$$c_{n}\sim{1\over2}\left\{\lambda(2n)\pm\sqrt{\lambda(2n)^{2}-4\epsilon
an}\right\}=\frac{\sqrt{aX}}{ sin(2z\sqrt{aX})}}\left[-1\pm cos(2z
\sqrt{aX})\right]_{z=2kn}=$$

$$-\sqrt{aX}
tan\left({{n\pi}\over{4}}+\sqrt{{{a{\epsilon}n}\over{2}}}+...
\right)\hskip.3cm and \hskip.3cm -\sqrt{aX}
cot\left({{n\pi}\over{4}}+\sqrt{{{a{\epsilon}n}\over{2}}}+...  \right).$$

{\bf Remark.} Behavior of C-type ("finite", see Introduction) solutions of
(4) for large but finite $n$ is characterized by formulae (42) as well, they
just happen to obtain zero meaning for some integer $N:\ c_{2N}=0\ or\
c_{2N+1}=0,\ N\in {\bf Z}$, which interrupts the lattice.

\par\vskip.5cm {\bf Appendix.} Proof of the Lemma.

\par Examine first the case (1) (Cauchy problem with i.d.
$\lambda_{1}(X_{0})=\lambda_{1}<\lambda_{2}(X_{0})=\lambda_{2}<
\lambda_{3}(X_{0})=0<\lambda_{4}(X_{0})=\lambda_{4}$)

$$\partial_{X}\lambda_{1}=-4a\frac{\Pi}{(\lambda_{4}-\lambda_{1})K}<0;$$

$$\partial_{X}\lambda_{2}=-4a\frac{\Pi-K}
{(\lambda_{4}-\lambda_{2})K}>0;\hskip.5cm \partial_{X}\lambda_{4}>0,$$
all the functions $\lambda_{j}(X)$ are monotonous for $\Pi/K>$1.

Suppose

$$\lambda_{2}=O(1),\hskip.5cm \lambda_{4}=O(1),\ \lambda_{1}=O(1),\
\partial_{X}\lambda_{1}<-{{4a}\over{\lambda_{4}-\lambda_{1}}}\Rightarrow
\lim_{X\rightarrow\infty}{\partial_{X}\lambda_{1}}\neq0,$$
what contradicts with assumption $\lambda_{1}=$O(1).

Suppose
$$\lambda_{2}=O(1),\ \lambda_{4}=O(1),\
\lambda_{1}\rightarrow-\infty\Rightarrow c^2\rightarrow0,\
\tilde{k^2}\rightarrow-\frac{\lambda_{2}} {\lambda_{4}-\lambda_{2}}
\Rightarrow $$

$$\partial_{X}\lambda_{1}=\frac{4a\Pi}{\lambda_{1}K}[1+o(1)]\Rightarrow
\lambda_{1}=O(X^{-1/2})\Rightarrow  c^2=O(X^{-1/2})\Rightarrow $$

$$\Pi-K=c^2\int\limits_0\limits^1\frac{x^2dx}
{\sqrt{G(x)}}=O(X^{-1/2})\Rightarrow \partial_{X}\lambda_{2}=O(X^{-1/2})$$
- contradiction with assumption $\lambda_{2}=O(1)$.

Therefore,

$$\lim_{X\rightarrow\infty}\lambda_{2}(X)=0$$.
So, one has at last: $\lambda_{2}\rightarrow 0,\ c^2={{\lambda_{2}}\over
{\lambda_{2}}}\rightarrow 0,\ \tilde k^2\rightarrow 0$

$$\Rightarrow \lim_{X\rightarrow\infty}\Pi/K=1\Rightarrow
\partial_{X}\lambda_{1}\sim-\frac{4a}{\lambda_{4}-\lambda_{1}},\
\partial_{X}\lambda_{4}\sim\frac{4a}{\lambda_{4}-\lambda_{1}}\Rightarrow$$

$$\partial_{X}(\lambda_{4}-\lambda_{1}){\sim}-\frac{8a}
{\lambda_{4}-\lambda_{1}}\Rightarrow\lambda_{1}(X)\sim-2\sqrt{aX},\
 \lambda_{4}(X){\sim}2\sqrt{aX}$$

Substituting this into the second eq. of (40), one gets the claim.  Part (2)
of the proof is quite similar.

\end{document}